\documentstyle[12pt]{article}

\newcommand{\dpar}[2]{\frac{\partial #1}{\partial #2}}
\newcommand{\vpar}[2]{\frac{\delta #1}{\delta #2}}
\newcommand{\jac}[2]{\frac{D(#1)}{D(#2)}}
\def\be{\begin{equation}}
\def\ee{\end{equation}}
\def\ltap{\raisebox{-.55ex}{\rlap{$\sim$}} \raisebox{.4ex}{$<$}}

\def\lsim{\mathrel{\ltap}}

\def\x{\mbox{\bf x}}
\begin{document}
\begin{flushright}
PURD-TH-94-11 \\
November 1994  \\
hep-ph/9411427
\end{flushright}
\vspace{0.5in}
\begin{center}
{\Large
The current of fermions scattered off a bubble wall } \\
\vspace{0.4in}
S.Yu. Khlebnikov\footnote{
Alfred P. Sloan Foundation Fellow; DOE Outstanding Junior Investigator}
\\
\vspace{0.1in}
{\it Department of Physics, Purdue University,
West Lafayette, IN 47907, USA} \\
\vspace{0.5in}
{\bf Abstract} \\
\end{center}
Proceeding from WKB quantization conditions, we derive a semiclassical
expression for the current of fermions scattered off a propagating
bubble wall in the presence of longitudinal gauge field. It agrees with
the expression used by Nasser and Turok in semiclassical analysis of
instability of electroweak bubble walls with respect to longitudinal
$Z$ condensation. We discuss the resulting dispersion relation for
longitudinal $Z$ field and show that light species are important for
the analysis of stability, because of their large contribution to
plasma frequency.
\newpage
\section{Introduction}
In a recent preprint \cite{NT}, Nasser and Turok have found that
interactions of bubble walls propagating during a first order electroweak
phase transition with fermions can lead to a $CP$ violating instability
on the walls, a condensate of longitudinal $Z$ bosons.
This result, if confirmed by comprehensive study, is very important
because it bears upon the main question in the theory of electroweak
baryogenesis, namely, whether successful electroweak baryogenesis requires
physics beyond the standard model or it does not.
Because of deviations from equilibrium that occur during the wall
propagation, the fermionic current, which
enters the equation of motion for the $Z$ field and contributes to its
effective mass, cannot be obtained from
thermodynamic relations and should be computed by direct averaging
of a microscopic expression.

As a first approach to the problem, Nasser and Turok considered
the WKB approximation for fermions and the thin-wall limit.
(They have also checked results against a fully quantum-mechanical solution.)
The expression for the current density they used in the WKB case would be
obviously correct if the local value of canonical momentum of fermion
remained unchanged during adiabatic switching on of the $Z$ field.
For a non-uniform Higgs field (bubble wall), this is not so, hence, a
justification for that expression is required.
In this note we shall show how that expression can be derived from
WKB quantization conditions for fermions via certain manipulations
with partial derivatives, similar to those used in thermodynamics.
We shall also discuss the resulting dispersion relation for longitudinal
$Z$ field and show that, in realistic case, not only the top quark but
also light species are important for the analysis of stability, because
of their large contribution to plasma frequency.

\section{Derivation of the formula for the current}
The microscopic expression for the current density
of a given species in the thin-wall case is
\be
J(\x) = \sum_n f_n j_n(\x) \; ,
\label{tot}
\ee
where $j_n$ is the current density for the $n$-th single-particle mode
and $f_n$ is the corresponding filling factor.
The use of single-particle modes is specific of the thin-wall
approximation that assumes that particles do not interact with each other
while scattering off the wall.
Though for a planar wall the non-trivial coordinate dependence is only
that for the direction orthogonal to the wall,
we keep the full three-dimensional
coordinate in (\ref{tot}) because some of the expressions below are
more general than that special situation.

The partial current densities $j_n$ are found by averaging the current density
operator over eigenfunctions of the corresponding Dirac equation.
If the Dirac hamiltonian is $H$, the current operator is
\be
j(\x)=-\vpar{H}{Z(\x)} \; .
\label{ope}
\ee
where $Z$ is some component of the field (in our case, that orthogonal to
the wall). The eigenfunctions $\psi_n$ satisfy
\be
H\psi_n = E_n \psi_n \; .
\label{eig}
\ee
By a variant of the Feynman-Hellmann theorem
(see for example ref.\cite{Griffiths})
\be
j_n(\x)=\psi_n^*  j(\x) \psi_n = -\vpar{E_n}{Z(\x)} \; .
\label{adi}
\ee
Hence, the partial currents are related to adiabatic, fixed $n$ variations of
the energy eigenvalues with respect to the field.

We can transform the fixed $n$ variations to fixed energy variations via
\be
\vpar{E_n}{Z(\x)}=-\left( \dpar{E_n}{n} \right)_Z
\left( \vpar{n}{Z(\x)} \right)_E \; ,
\label{ene}
\ee
which is derived by using jacobians (cf. ref.\cite{LL})
\be
\jac{E,n}{Z,n}=\jac{E,n}{E,Z} ~\jac{E,Z}{Z,n} \; .
\label{jac}
\ee
The resulting expression for the current is quite general (once the
thin-wall limit is assumed) but it
is especially useful in the WKB approximation, because in that case
the quantum numbers have simple representation through WKB quantization
conditions.

Like ref.\cite{NT}, we assume the potential barrier created for particles
by the wall and the $Z$ field together to be monotonic.
The WKB quantization conditions in the direction orthogonal to the wall
(the only non-trivial ones) are
\begin{eqnarray}
\int_{-L_z/2}^{L_z/2} p_z(z) dz & = & 2\pi n_z \; ,
\label{qua1} \\
2 \int_{-L_z/2}^{z^*(n)} p_z(z) dz & = & 2\pi n_z + C
\label{qua2}
\end{eqnarray}
for transmitted and reflected fermions, respectively.
Here $p_z(z)$ is the longitudinal component of fermion momentum at a
given spatial point, as determined from the classical
dispersion law by conservation of single-particle energy and transverse
momentum. $n_z$ is a positive or negative integer.
The classical turning point $z^*$ of reflected particles depends on
state $n=(n_x,n_y,n_z)$.
A constant (up to exponential in $-L_z$ terms) phase shift $C$ depends on
the boundary condition for reflected particles at $-L_z/2$.
The dispersion laws are listed in ref.\cite{NT}.

A single-particle mode can be specified by the value of momentum at one of
the infinities (more precisely, at $\pm L_z/2$), $\mbox{\bf p}_{\infty}$.
That incident momentum is related to the energy of the mode by the usual,
field-independent expression. So, at given $p_{x,\infty}$, $p_{y,\infty}$,
the variation at fixed energy is variation at fixed $p_{z,\infty}$.

Next, we use $n_z$ instead of $n$ in (\ref{ene}).
Going over to one-dimensional variations we obtain
\be
\vpar{E_n}{Z(z)}=-\frac{1}{L_x L_y}\left( \dpar{E_n}{n_z} \right)_Z
\left( \vpar{n_z}{Z(z)} \right)_{p_{z,\infty}}
=-\frac{\eta}{(2\pi)L_x L_y} \left( \dpar{E_n}{n_z} \right)_Z
\left( \dpar{p_z(z)}{Z(z)} \right)_{p_{z,\infty}}
\label{one}
\ee
where $\eta$ is 1 for transmitted states, 0 for reflected states with
$z^*<z$, and 2 for reflected states with $z^*>z$. Note that the variational
derivative with respect to $Z$ is converted into ordinary one.
The derivatives are
taken at fixed $p_{x,\infty}$, $p_{y,\infty}$, which is not indicated
explicitly in (\ref{one}).

Using (\ref{one}) and another transformation of derivatives
\be
\left( \dpar{p_z(z)}{Z(z)} \right)_{p_{z,\infty}}
=- \left( \dpar{p_z(z)}{p_{z,\infty}} \right)_{Z}
   \left( \dpar{p_{z,\infty}}{Z} \right)_{p_z} \; ,
\label{ano}
\ee
we transform the current density as follows
\begin{eqnarray}
J(z) & = & \int dn_xdn_ydn_z f_n j_n(z) \nonumber \\
 & = & - \int \eta \frac{dp_{x,\infty}dp_{y,\infty}}{(2\pi)^3} dn_z
f(\mbox{\bf p}_{\infty})
\left( \dpar{E_n}{n_z} \right)_Z
\left( \dpar{p_z(z)}{p_{z,\infty}} \right)_{Z}
\left( \dpar{p_{z,\infty}}{Z} \right)_{p_z} \nonumber \\
 & = & - \int \eta \frac{dp_{x,\infty}dp_{y,\infty}}{(2\pi)^3} dn_z
f(\mbox{\bf p}_{\infty})
\dpar{E}{p_{z,\infty}}
\left( \dpar{p_{z,\infty}}{n_z} \right)_Z
\left( \dpar{p_z(z)}{p_{z,\infty}} \right)_{Z}
\left( \dpar{p_{z,\infty}}{Z} \right)_{p_z} \nonumber \\
 & = & - \int \eta \frac{d^3p_{\infty}}{(2\pi)^3}
f(\mbox{\bf p}_{\infty})
\left( \dpar{p_z(z)}{p_{z,\infty}} \right)_{Z}
\left( \dpar{E}{Z} \right)_{p_z} \; .
\label{fin}
\end{eqnarray}
The last line is the expression used in ref.\cite{NT}.
We have thus shown how it can be derived from WKB quantization conditions.

\section{Dispersion relation and plasma frequency}
The formula for the current allows us to compute
the response of the non-equilibrium plasma to longitudinal
$Z$ field. Because in the WKB calculation of the current both the wall
profile and the gauge field were assumed slowly varying, we are
actually considering the zero-momentum limit of the response.
Accordingly, the relation (\ref{fin}) between the current and the field
is local in space.

The linearized equation of motion --- the dispersion relation for
the $Z$ field is
\be
\mbox{diag}(\omega^2  - m^2_Z(z),~\omega^2) - {\hat \omega}_p^2(z) = 0 \; .
\label{dis}
\ee
This is a matrix equation because of the mixing between $Z$ and photon
field $A$ due to plasma effects;
${\hat \omega}_p^2(z)$ is the non-equilibrium plasma frequency matrix,
\be
{\hat \omega}^2_p(z) =
-\left( \begin{array}{rr}
\partial J(z)/\partial Z(z) & \partial J(z)/\partial A(z) \\
\partial J_{em}(z)/\partial Z(z) & \partial J_{em}(z)/\partial A(z)
\end{array} \right)_{Z=A=0} \; ,
\label{ome}
\ee
where $J$ now denotes the total, summed over all species, current coupled
to $Z$ and $J_{em}$ is the total electromagnetic current.

Non-equilibrium effects in the plasma frequency are important only for
the heaviest fermion, the top quark, which interacts effectively with the
bubble wall. Evaluation of the integral (\ref{fin}) shows \cite{NT}
that for non-zero wall velocity, fermions produce
a contribution to the current which has a square-root singularity in
$m_{\infty}-m(z)$, where $m_{\infty}$ is the mass of a fermion
in the broken phase at infinity and $m(z)$ is its mass locally
(so the singularity is behind the wall).
This singular term corresponds to a negative contribution
to the upper-left entry of the plasma frequency matrix,
\be
\Delta \omega^2_{p11}(z) \sim - \frac{\alpha_W}{4\pi\cos^2\theta_W}
\frac{u m_{\infty}^3}{[m^2_{\infty}-m^2(z)]^{1/2}}  \; ,
\label{sin}
\ee
where $\alpha_W=g^2/(4\pi)$ is the weak interaction constant,
$u$ is the wall velocity.
One can see that the typical value of $p_z(z)$ that give rise
to the singular term (\ref{sin}) is of order $[m_{\infty}-m(z)]^{1/2}$.
For the WKB approximation to be applicable, this should be large compared to
the inverse thickness of the wall given roughly by the Higgs mass $m_H$
(all masses are those at the phase transition temperature).
So, within the WKB domain, the singularity provides an enhancement factor
of order
\be
m_{\infty}/[m_{\infty}^2 - m^2(z)]^{1/2} \lsim m_{\infty}/m_H \; ,
\label{enh}
\ee
which can be considerable for top quarks.

However, to decide if non-equilibrium contribution of top quarks to
the dispersion relation, whether obtained by WKB or fully
quantum-mechanical means, leads to an instability of the $Z=0$ state,
we should consider it together with contributions of other
particles.\footnote{The author is grateful to M. Shaposhnikov for
discussion that led to these considerations.}
In particular, numerous light species (those with masses much smaller
than temperature) give an essentially equilibrium contribution
to ${\hat \omega}_p^2$.
The equilibrium contribution of fermions and {\em transverse} $W$ bosons
to ${\hat \omega}_p^2$ can be obtained simply by orthogonal transformation
of the diagonal matrix of plasma frequencies of the $SU(2)\times U(1)$
basis. With $N_g$ light fermionic generations, light transverse $W$, and
small wall velocity, this contribution is
\be
({\hat \omega}_p^2)_{eq} \approx \frac{g^2 T^2}{9c^2}
\left( \begin{array}{rr}
N_g \left( 1-2s^2+\frac{8}{3}s^4\right) + 2c^4 &
N_g \left( 1-\frac{8}{3}s^2\right)sc  + 2sc^3 \\
N_g \left( 1-\frac{8}{3}s^2\right)sc + 2sc^3 &
\left( \frac{8}{3} N_g + 2 \right) s^2 c^2
\end{array} \right) \; ,
\label{equ}
\ee
where $c\equiv\cos\theta_W$, $s\equiv\sin\theta_W$.
For our estimates, we took $N_g=3$ but subtracted the top-quark contribution
from (\ref{equ}), considered
$W$ bosons as light, and neglected the role of longitudinal $W$ and
Higgs bosons.

In the WKB approximation, the quantity of interest is the magnitude of the
non-equilibrium top-quark contribution to ${\hat \omega}_p^2$, the most
significant part of which is (\ref{sin}), corresponding to the onset of
instability. It is determined approximately from
\be
\det \left( \begin{array}{cc}
\Delta \omega^2_{p11}(z) + m^2_Z(z) + A  &  ~~~~~B \\
B & ~~~~~C \end{array} \right) = 0 \; ,
\label{det}
\ee
where $A$, $B$, $C$ are the entries of the equilibrium plasma frequency
matrix of light species.
(In a fully quantum-mechanical, non-WKB calculation, the relation between
the current and the field becomes non-local \cite{NT} and so does
the criterion for instability.)
Estimating the plasma frequency of light species as described above, we find
from (\ref{det}) that the onset of instability is at
\be
- \Delta \omega^2_{p11}(z) =  m^2_Z(z) + 0.25 g^2 T^2 /\cos^2\theta_W \; .
\label{ons}
\ee
For realistic values of $\phi_{\infty}/T$, where $\phi_{\infty}$ is the
expectation value of the Higgs field in the broken phase,
the second term on the right-hand
side is of the same order as $m_Z^2$ in the broken phase, for example,
for $\phi_{\infty}=1.5 T$ used in ref.\cite{NT}, about 45\% of it.
We conclude that light species are in general important
for the analysis of stability, because of their large contribution to
the dispersion relation.

The author is grateful to N. Turok for a lot of useful correspondence
and sending parts of the longer paper prior to publication,
and to T. Clark, G. Giuliani, S. Love and M. Shaposhnikov for discussions.
This work was supported in part by the Alfred P. Sloan Foundation
and in part by the U.S. Department of Energy.

\end{document}